\documentclass[PRApplied, twocolumn, amsmath,amssymb, aps]{revtex4-1} 
\usepackage{graphicx} 
\usepackage{hyperref} 
\usepackage{color}
\usepackage[yyyymmdd,hhmmss]{datetime}
\usepackage[normalem]{ulem}
\usepackage{todonotes}

\begin{document}
\title{Electrical control of spin mixing conductance in a Y${}_{3}$Fe${}_{5}$O${}_{12}$/Platinum bilayer}
\author{Ledong Wang} 
\affiliation{School of Physics, State Key Laboratory of Crystal Materials, Shandong University, 27 Shandanan Road, Jinan, 250100 China}
\author{Zhijian Lu} 
\affiliation{School of Physics, State Key Laboratory of Crystal Materials, Shandong University, 27 Shandanan Road, Jinan, 250100 China}
\author{Jianshu Xue} 
\affiliation{School of Physics, State Key Laboratory of Crystal Materials, Shandong University, 27 Shandanan Road, Jinan, 250100 China}
\author{Peng Shi} 
\affiliation{School of Physics, State Key Laboratory of Crystal Materials, Shandong University, 27 Shandanan Road, Jinan, 250100 China}
\author{Yufeng Tian} 
\affiliation{School of Physics, State Key Laboratory of Crystal Materials, Shandong University, 27 Shandanan Road, Jinan, 250100 China}
\author{Yanxue Chen} 
\affiliation{School of Physics, State Key Laboratory of Crystal Materials, Shandong University, 27 Shandanan Road, Jinan, 250100 China}
\author{Shishen Yan}
\email{shishenyan@sdu.edu.cn}
\affiliation{School of Physics, State Key Laboratory of Crystal Materials, Shandong University, 27 Shandanan Road, Jinan, 250100 China}
\author{Lihui Bai} 
\email{lhbai@sdu.edu.cn}
\affiliation{School of Physics, State Key Laboratory of Crystal Materials, Shandong University, 27 Shandanan Road, Jinan, 250100 China}
\author{Michael Harder}  
\affiliation{Department of Physics, Kwantlen Polytechnic University, 12666 72 Avenue, Surrey, BC V3W 2M8 Canada}
\date{\today/\currenttime}

\begin{abstract} 
We report a tunable spin mixing conductance, up to $\pm 22\%$, in a Y${}_{3}$Fe${}_{5}$O${}_{12}$/Platinum (YIG/Pt) bilayer.  
This control is achieved by applying a gate voltage with an ionic gate technique, which exhibits a gate-dependent ferromagnetic resonance line width.
Furthermore, we observed a gate-dependent spin pumping and spin Hall angle in the Pt layer, which is also tunable up to $\pm$ 13.6\%.
This work experimentally demonstrates spin current control through spin pumping and a gate voltage in a YIG/Pt bilayer, demonstrating the crucial role of the interfacial charge density for the spin transport properties in  magnetic insulator/heavy metal bilayers.
\end{abstract}

\maketitle
\section{introduction}
Spin currents in Y${}_{3}$Fe${}_{5}$O${}_{12}$/Platinum (YIG/Pt) bilayers have attracted much attention in the past decade due to the unique spin current transport properties in magnetic insulators and spin charge conversion in heavy metals \cite{Kajiwara2010Nature, Lu2012PRL, Bai2013PRL, Hahn2013PRB, Sun2013PRL, Castel2014PRB, Hyde2014PRB,  Haertinger2015PRB, Zhou2016PRB,  Dushenko2016PRL, Wang2016PRL, Wesenberg2017NatPhysics,  Kapelrud2017PRB, Keller2017PRB}.
Mechanisms including spin pumping at the bilayer interface \cite{Tserkovnyak2002PRL}, spin diffusion \cite{Montoya2014PRL}  and the inverse spin Hall effect in heavy metals \cite{Saitoh2006APL}, have been established to interpret the generation, transfer and conversion of spin current \cite{Tserkovnyak2005RMP}.  This understanding has advanced the development of spintronic devices, as evidenced by, for example, nano oscillators.  
However  spintronic devices based on spin pumping require additional control beyond the basic generation and detection of spin current \cite{Igor2004RMP}. 
In this regard improvement of spin current transport at the interface of the YIG/Pt bilayer is the kernel toward the application of spin current due to spin pumping.

The efficiency with which the spin current crosses a bilayer interface is characterized by the spin mixing conductance, $g_{\uparrow\downarrow}$, which is a constant for a given sample and is usually measured by  comparing the ferromagnetic resonance (FMR) line width in a bilayer device to that of a bare YIG layer.
Some attempts have been made to change this interfacial spin transport. 
For example, a thin NiO \cite{Du2013PRL, Wang2014PRL} or CrO \cite{Qiu2018NatMaterials} layer inserted at the interface has been reported to enhance or suppress the spin current, respectively. 
Additionally, Pt alloyed with Al or Au was demonstrated to enhance the spin transfer torque at the interface \cite{Nguyen2016APL}. 
Importantly, such previous works hint that the conductivity in the Pt layer may change the boundary at the interface, hence changing the spin transport properties as well as the spin mixing conductance.

\begin{figure}[b]
    \includegraphics[width = 8.17 cm]{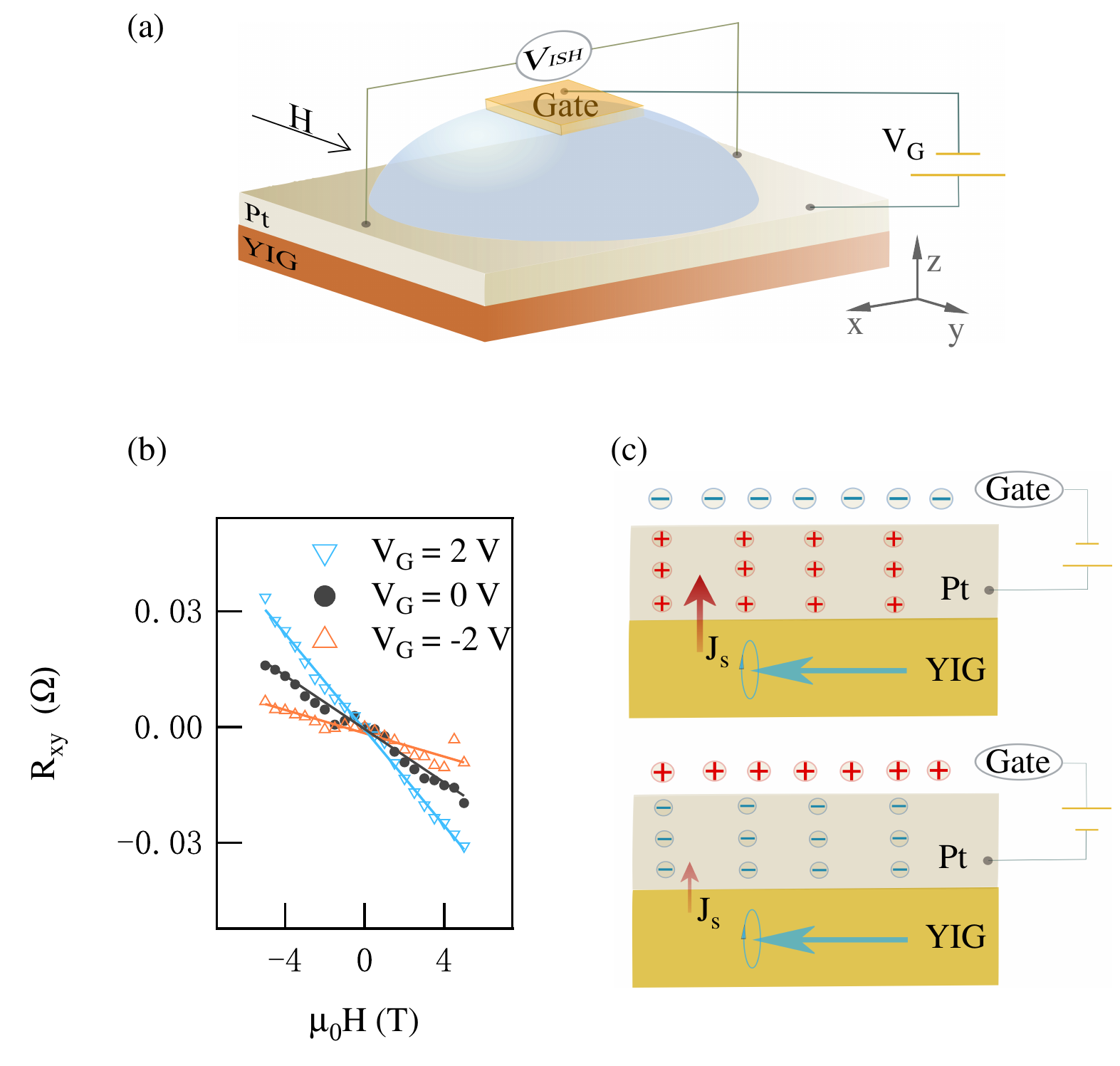}
    \caption{(a) Spin pumping measurement setup using a YIG/Pt bilayer with a gate voltage $V_{G}$ applied using an ionic gate technique. An external magnetic field {\bf H} was applied along the $y$-axis perpendicular to the measurement direction ($x$-axis). (b) A Hall measurement in a 3-nm-thick Pt film indicates that the carrier density was modulated by the gate voltage.   (c)  Sketch of the carrier density change due to the gate voltage, which changes the boundary conditions at the YIG/Pt interface, influencing the spin current.}
    \label{fig1}
\end{figure}

To control the spin transport properties of bilayer samples a gate voltage can be used to control interfacial boundary conditions.
 Interestingly gate voltage techniques, originally developed to control the carrier density in semiconductors, have recently been applied to thin metal films \cite{Wang2012NatMaterials}. 
In this context the addition of an ionic gate has been shown to enlarge the field effect in metallic films due to the huge number of ions accumulated at the interface under certain bias voltage conditions \cite{Yamada2011Science}.
As a result it has been reported that the carrier density and anomalous Hall effect in Pt may be modulated \cite{Shimizu2013PRL, Liang2018PRB}.
This brings about the interesting possibility to develop additional control techniques to manipulate spin transport at the YIG/Pt interface \cite{Guan2018AM}. 

In this work we applied a gate voltage to a YIG/Pt bilayer using an ionic gate technique to modulate the charge accumulation at the bilayer interface. 
We experimentally observed that the FMR line width is both enhanced and suppressed depending on the polarity of the gate voltage.
By compare the FMR line width in the bilayer to that of the bare YIG thin film, we found a gate tunable spin mixing conductance in the YIG/Pt bilayer. 
Additionally, the observation of a shift in the FMR resonance field towards both high and low fields, depending on the polarity of the gate voltage, allows us to rule out the Joule heating effect.
This shift in the FMR resonance field indicates that the effective field of the magnetization was changed by the gate voltage which may be induced by charge accumulation at the interface.   
Using the Landau-Lifshitz-Gilbert equation and spin pumping theory, we evaluated the spin current and the spin Hall effect in Pt, finding a strong gate voltage dependence.
Thus we have experimentally demonstrated control of the spin current transport at a bilayer interface, which will be useful for understanding spin transfer at the interfaces of magnetic insulator and heavy metal bilayers.  
 
\section{EXPERIMENTAL METHODS}
To fabricate the YIG/Pt bilayer a 20-nm thick YIG layer was first deposited on a GGG substrate using pulsed laser deposition.
The YIG had a saturation magnetization of $\mu_{0}M_{s}$ = 0.175 T and a Gilbert damping of $\alpha_{YIG}$ = 0.00049.
A 2.5-nm-thick Pt layer was then deposited on top of the YIG using magneto sputtering in which the base pressure and the sputtering pressure were $2\times 10^{-5}$ Pa and 0.68 Pa, respectively. 
The Gilbert damping of the bilayer was $\alpha_{YIG/Pt}$ = 0.00157 which is 3 times larger than that of the bare YIG film. 
The lateral dimensions of the bilayer were 5 mm $\times$ 2.5 mm, and a Hall bar was fabricated using lithography techniques for reference and Hall measurements.
 
To measure the spin current signal due to spin pumping the YIG/Pt  bilayer was driven to FMR by microwaves applied through a coplanar waveguide beneath the sample.
The microwave output power was 158 mW and the modulation frequency of the microwave power (used for lock-in measurements) was 8.33 kHz.
An external magnetic field {\bf H} was applied in-plane and perpendicular to the Pt strip to enhance the spin pumping signal. 
The spin pumping voltage was measured along the $x$-axis of the Pt layer using a lock-in technique while sweeping the magnetic field {\bf H} at room temperature.

An ionic gate, composed of a composite solid electrolyte, was placed on top of the Pt layer and a gate voltage $V_{G}$ was applied between a contact (labelled as Gate) on top of the solid electrolyte, and the Pt layer. 
The composite solid electrolyte used for gating was prepared using 81 wt$\%$ of acrylic resin, 14.6 wt$\%$ of succinonitrile, and 4.4 wt$\%$ of cesium perchlorate.
As shown in Fig. \ref{fig1} (b) the sheet carrier density in the 3-nm-thick Pt layer was found to be 1.8 $\times 10^{17}$ cm$^{-2}$ when no gate voltage was applied, $V_{G}$ = 0 V, assuming the single band relation $R_{H} =  1/ne$ ($e$ is the electron charge and $n$ is the carrier density).
This is comparable to the reported results for Pt \cite{ Shimizu2013PRL}.
However, as indicated by the squares and triangles for $V_{G}$ = 2 V and -2 V respectively, the Hall resistance, and hence the carrier density, changes significantly when a gate voltage is applied.
For a 3-nm-thick Pt thin film it is expected that the boundary at the interface of the YIG and Pt layer is strongly dependent on the carrier density as well as the spin orbit interaction in the Pt layer as shown in Fig. \ref{fig1} (c). 
Thus the change in boundary conditions due to charge accumulation at the YIG/Pt interface will change the interfacial spin transport properties, which was experimentally observed in the FMR measurement. 
In our work we defined a positive gate voltage when the carrier density was reduced and vice versa. 
As a reference the FMR absorption in a bare YIG film was also measured as a function of the gate voltage\cite{SuppleM}.  In this case the voltage induced FMR changes were found to be small and ignorable.

\section{RESULTS AND DISCUSSION}

 \subsection{$V_{G}$-dependent spin mixing conductance}
\begin{figure}[tb]
    \includegraphics[width =8.17 cm]{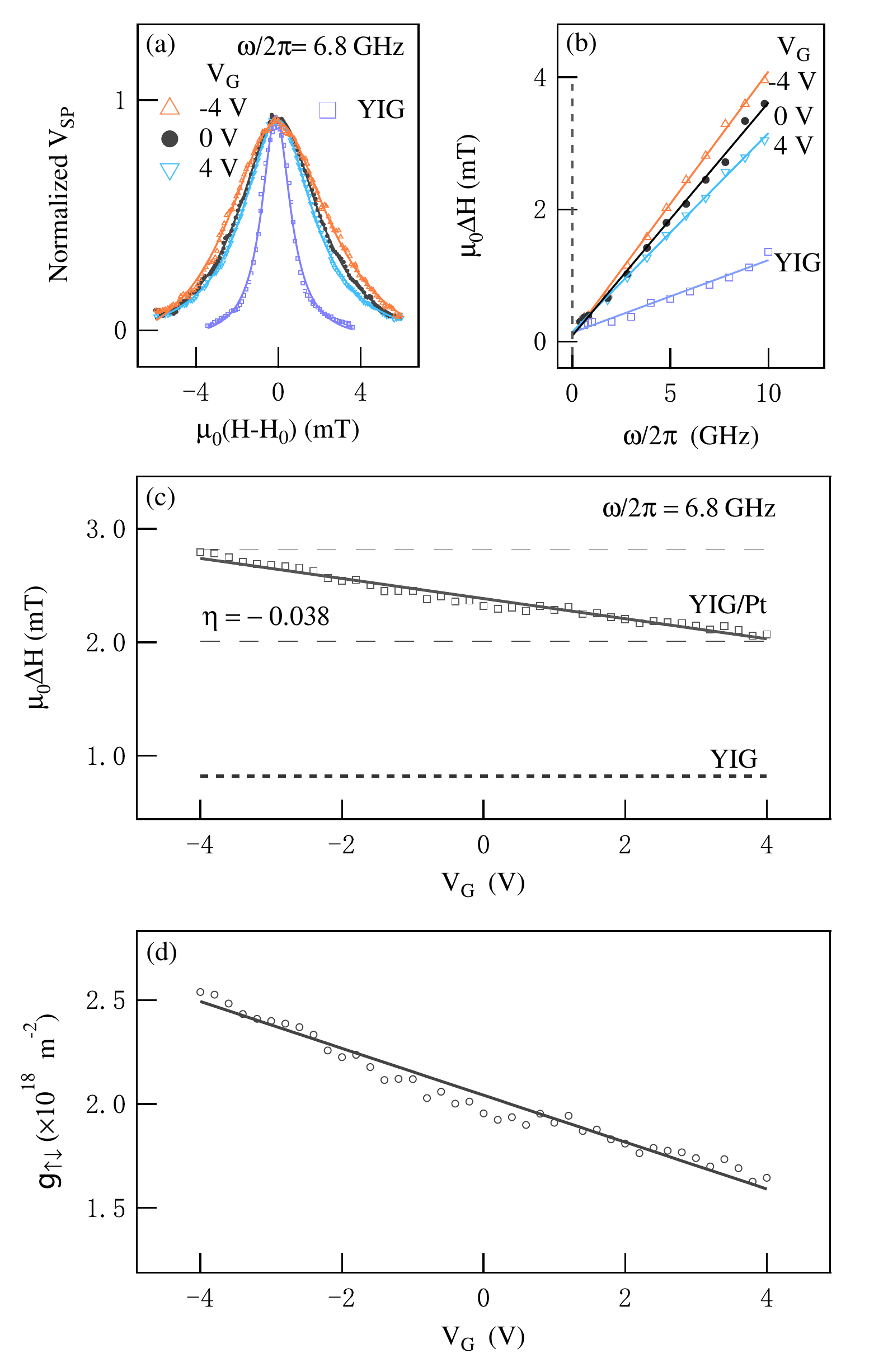}
    \caption{Spin pumping voltage in the YIG/Pt bilayer, normalized by the maximum amplitude to highlight the line width changes induced by the gate voltage $V_{G}$. The FMR absorption signal (squares) from a bare YIG layer is used as a reference. (b) The FMR line width as a function of microwave frequency depends on the gate voltage.  (c) The FMR line width $\Delta$H is plotted as a function of the gate voltage $V_{G}$ and compared to that in bare YIG. The solid line indicates the fitting results to Eq. \eqref{eq_linewidth}. Here the microwave frequency is 6.8 GHz. (d)  Spin mixing conductance is subtracted and predicted based on Eqs. \eqref{eq_spinmixingconductance} and \eqref{eq_linewidth}.  }
    \label{fig2}
\end{figure}
 
Figure \ref{fig2} (a) shows the spin pumping voltage measured in the YIG/Pt bilayer for a variable gate voltage $V_{G}$, with the FMR absorption signal in the YIG bare layer plotted for reference.  
Here the FMR spectra has been shifted by the resonance field $H_{0}$ and normalized to the maximum signal amplitude, in order to highlight the FMR line width change due to the gate voltage. 
The significant broadening of the YIG/Pt line width, as compared to the FMR line width in the bare YIG layer,  has been experimentally observed and theoretically studied previously \cite{Weiler2013PRL, Tserkovnyak2002PRL}.
Here we also observe that the bilayer line width is dependent on the applied gate voltage $V_{G}$.
Figure \ref{fig2} (b) shows the FMR  line width $\Delta H$ (half-width-half-maximum) in both the YIG and YIG/Pt samples. 
The Gilbert damping, determined from the gradient of the line width as a function of the microwave frequency, is enhanced in the YIG/Pt bilayer compared to that in YIG.
Clearly the gate voltage applied to the YIG/Pt bilayer suppresses and enhances the Gilbert damping of the bilayer. 
The inhomogeneous broadening $\Delta H_{0}$  is around  0.1 mT which is one order smaller than the line width $\Delta H$ at 6.8 GHz in both the YIG/Pt and YIG samples. 
Furthermore $\Delta H_{0}$  was barely affected by the gate voltage. 
Therefore it is a good approximation to subtract the Gilbert damping directly using the line with $\Delta H$ at 6.8 GHz.
Such a line width change as a function of $V_{G}$ is summarized in Fig. \ref{fig2} (c), where the dashed horizontal line indicates the FMR line width of the bare YIG. 
In a first order approximation we assume that the line width $\Delta H$ is influenced by the gate voltage $V_{G}$, according to

\begin{equation}
\Delta H = \Delta H_{0}+ \frac{\omega}{\gamma} \alpha_{YIG/Pt}(1 + \eta V_{G}).
\label{eq_linewidth}
\end{equation}
Here $\Delta H_{0}$ is the frequency independent inhomogeneous broadening,  the gyromagnetic ratio $\gamma = 2\pi \mu_{0} \times $28 GHz/T, the Land\'e factor $g =2$, $\mu_{B}$ is the Bohr magneton, $\omega$ is microwave angular frequency, $ \alpha_{YIG/Pt}$ is the Gilbert damping of the YIG/Pt bilayer and $\eta$ is a proportionality factor that characterizes the influence of the gate voltage on the Gilbert damping.  The units of $\eta$ are $V^{-1}$.
Here we find $\eta$  = -0.038 $V^{-1}$ by fitting the line width of the YIG/Pt bilayer to Eq. \eqref{eq_linewidth}.
Such a $V_{G}$-dependent line width was also observed for various different frequencies (not shown here). 
The spin mixing conductance $g_{\uparrow\downarrow}$ was experimentally evaluated by comparing the FMR line width $\Delta H_{YIG/Pt}$ in the YIG/Pt bilayer to the FMR line width $\Delta H_{YIG}$ in the bare YIG layer \cite{Weiler2013PRL}, 
\begin{equation}
g_{\uparrow\downarrow} = \frac{4\pi M_{s}\gamma t_{YIG}}{g\mu_{B}\omega}(\Delta H_{YIG/Pt} - \Delta H_{YIG}),
\label{eq_spinmixingconductance}
\end{equation}
where $M_{s}$ is the saturation magnetization and $t_{YIG}$ is the thickness of the YIG layer. 

By comparing the line width of FMR in the YIG/Pt bilayer to that in the YIG thin film according to Eq. \eqref{eq_spinmixingconductance}, one can evaluate the spin mixing conductance $g_{\uparrow\downarrow}$ as summarized in Fig. \ref{fig2} (d).
We find that $g_{\uparrow\downarrow}$ is roughly linearly-dependent on the gate voltage $V_{G}$ as indicated by the solid line, which can be predicted by Eq. \eqref{eq_spinmixingconductance} using $\eta$  = -0.038 $V^{-1}$.
This indicates that we have experimentally manipulated the YIG/Pt interfacial spin transport properties, which is a key issue concerning spin injection in the spintronics community.  We note that here the spin mixing conductance is an effective value since spin back flow will play a role in the 2.5-nm-thick Pt \cite{Jiao2013PRL, Du2014PRApplied}.

 \subsection{Spin current manipulated by $V_{G}$}

\begin{figure}[bt]
    \includegraphics[width =8.17 cm]{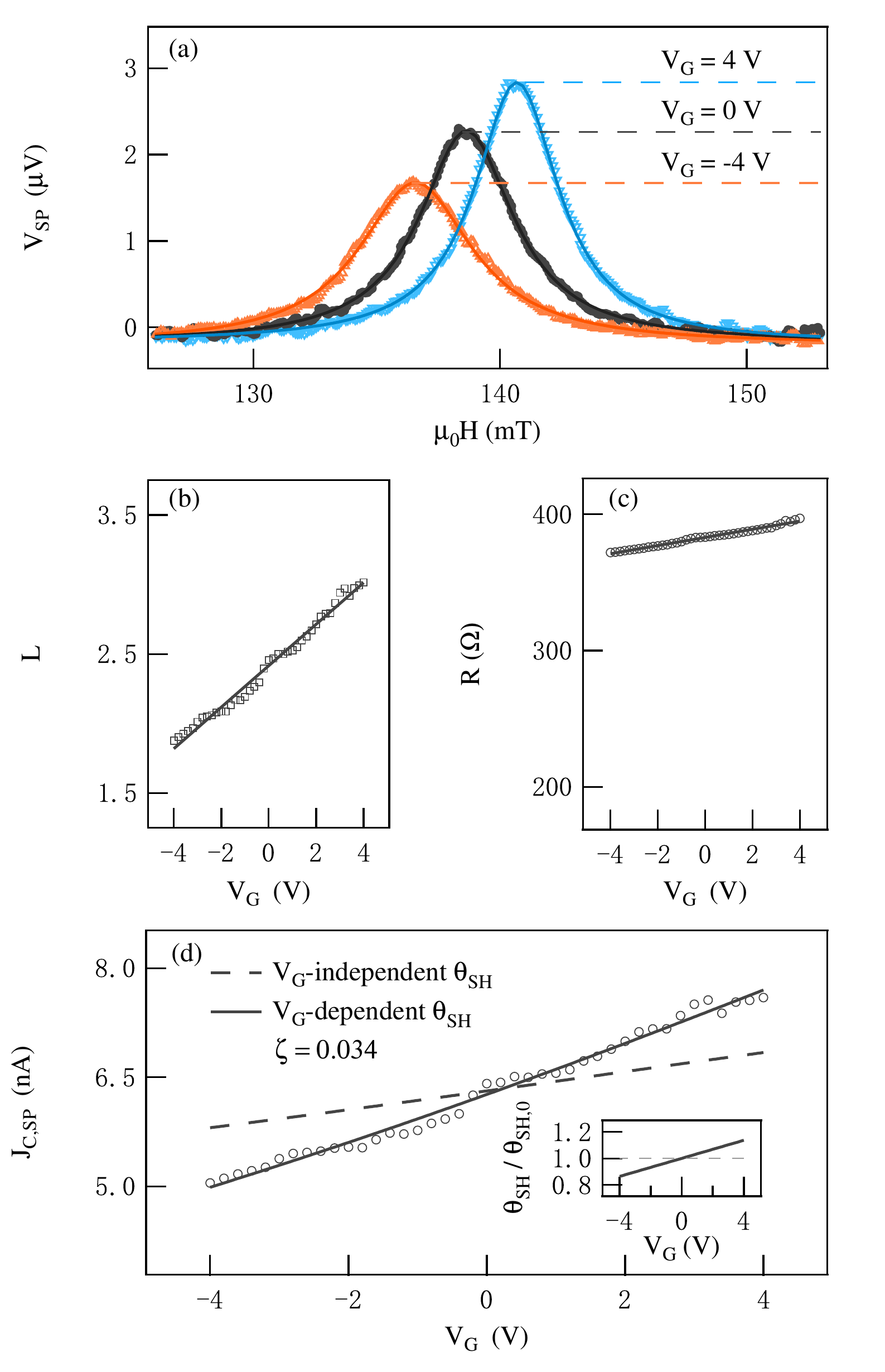}
    \caption{(a) Spin pumping voltages $V_{SP}$ at various $V_G$, demonstrating a gate voltage dependence. The spin pumping voltage $V_{SP}$ and the resistance of the YIG/Pt bilayer are plotted in (b) and (c) respectively as a function of the gate voltage $V_{G}$. (d) The charge current due to spin pumping $J_{C, SP}$ was compared to the predictions with and without the $V_{G}$-dependent spin Hall angle. The inset displays the spin Hall angle as a function of the gate voltage.  Here, the microwave frequency is 6.8 GHz.}
    \label{fig3}
\end{figure}
 
Since the spin mixing conductance is $V_{G}$ tunable we may expect that the spin current due to spin pumping is also controlled by the gate voltage.
As evidenced by the broadened line width, the gate voltage enhanced spin mixing conductance is the key source of additional FMR damping.  This leads to a reduced FMR amplitude and thus a (1/$\alpha$)${}^{2}$ decrease in the spin current pumped by FMR. Therefore even though the enhanced spin mixing conductance leads to a large spin current transparency at the bilayer interface, the observed spin current amplitude is reduced.
Figure \ref{fig3} (a) shows the spin pumping voltage at different values of $V_{G}$.
The amplitude of $V_{SP}$ is enhanced by applying a positive voltage and suppressed by a negative voltage, with a total change up to $\pm$46.3\% as shown in Fig. \ref{fig3} (b).
We also carefully examined the resistance change of the Pt layer as a function of $V_{G}$, which shows a much smaller change ($\pm$3.4\%) as highlighted in Fig. \ref{fig3} (c). 
Fig. \ref{fig3} (d) displays the $V_{G}$-dependent charge current  $J_{C, SP}$ which has been generated from the spin current through the inverse spin Hall effect of the Pt layer. 
The gate voltage dependence of the charge current can be evaluated by considering the Polder tensor for FMR in the YIG layer, the spin mixing conductance at the interface, the spin diffusion into the Pt layer, and the inverse spin Hall effect in the Pt layer.
For a given microwave frequency and power the spin current produced by the FMR and injected into the Pt layer at the interface will have the form $j_{s,0} \propto g_{\uparrow\downarrow}/\alpha_{YIG/Pt}^{2}$ and therefore the charge current may be written as, 
\begin{equation} 
J_{C, SP} = k\theta_{SH}g_{\uparrow\downarrow}/\alpha_{YIG/Pt}^{2}. \label{eq_chargecurrent}   
\end{equation}
Here $k$ is a $V_{G}$-independent constant which depends on the microwave frequency and power, saturation magnetization of the YIG and the spin current diffusion properties of the Pt layer.
Based on this analysis we can predict the $V_{G}$-dependent charge current using Eq. \eqref{eq_chargecurrent} as shown by the dashed line in Fig. \ref{fig3} (d), where $k\theta_{SH} =  7.64 \times 10^{-24}$ $nA\cdot m^{2}$. 
Although the predicted dashed line does have a $V_{G}$-dependence (due to the $V_G$-dependent spin mixing conductance), it does not match the experimental observation.
This indicates that for a given spin Hall angle, although the spin mixing conductance was enhanced, less spin current was produced than that required to generate the necessary charge current.
Therefore, in order to compare with the observed $J_{C, SP}$ it is reasonable to assume that the spin Hall angle, $\theta_{SH}$ in Eq. \eqref{eq_chargecurrent}, is also $V_{G}$-dependent,

\begin{equation} 
\theta_{SH} = \theta_{SH,0} (1 + \zeta V_{G}). \label{eq_spinhallangle}
\end{equation}
Here $\zeta$ defines the gate voltage dependence of the spin Hall angle and has units of $V^{-1}$ and
$ \theta_{SH,0}$ is the spin Hall angle for $V_{G}$ =  0 V.
By combining Eqs. \eqref{eq_chargecurrent} and \eqref{eq_spinhallangle}, we predict the charge current due to spin pumping as shown by the solid line in Fig. \ref{fig3} (b), which matches well with the experimental data.
In this calculation $k\theta_{SH,0} =  7.58 \times 10^{-23}$ $nA\cdot m^{2}$, which is the same as  $k\theta_{SH}$ used for Eq. \eqref{eq_chargecurrent} (dashed line), and $\zeta$ = 0.034 $V^{-1}$ indicating that the spin Hall angle in the Pt layer is also tunable by the gate voltage.
Therefore, we find a spin Hall angle which can be tuned by up to $\pm$13.6\% as shown by the inset in Fig. \ref{fig3} (d).  This tendency is opposite the behaviour observed for the $V_{G}$-dependent spin mixing conductance.
 
 \subsection{$V_{G}$-dependent anisotropy field} 
 
\begin{figure}[tb]
    \includegraphics[width =8.17 cm]{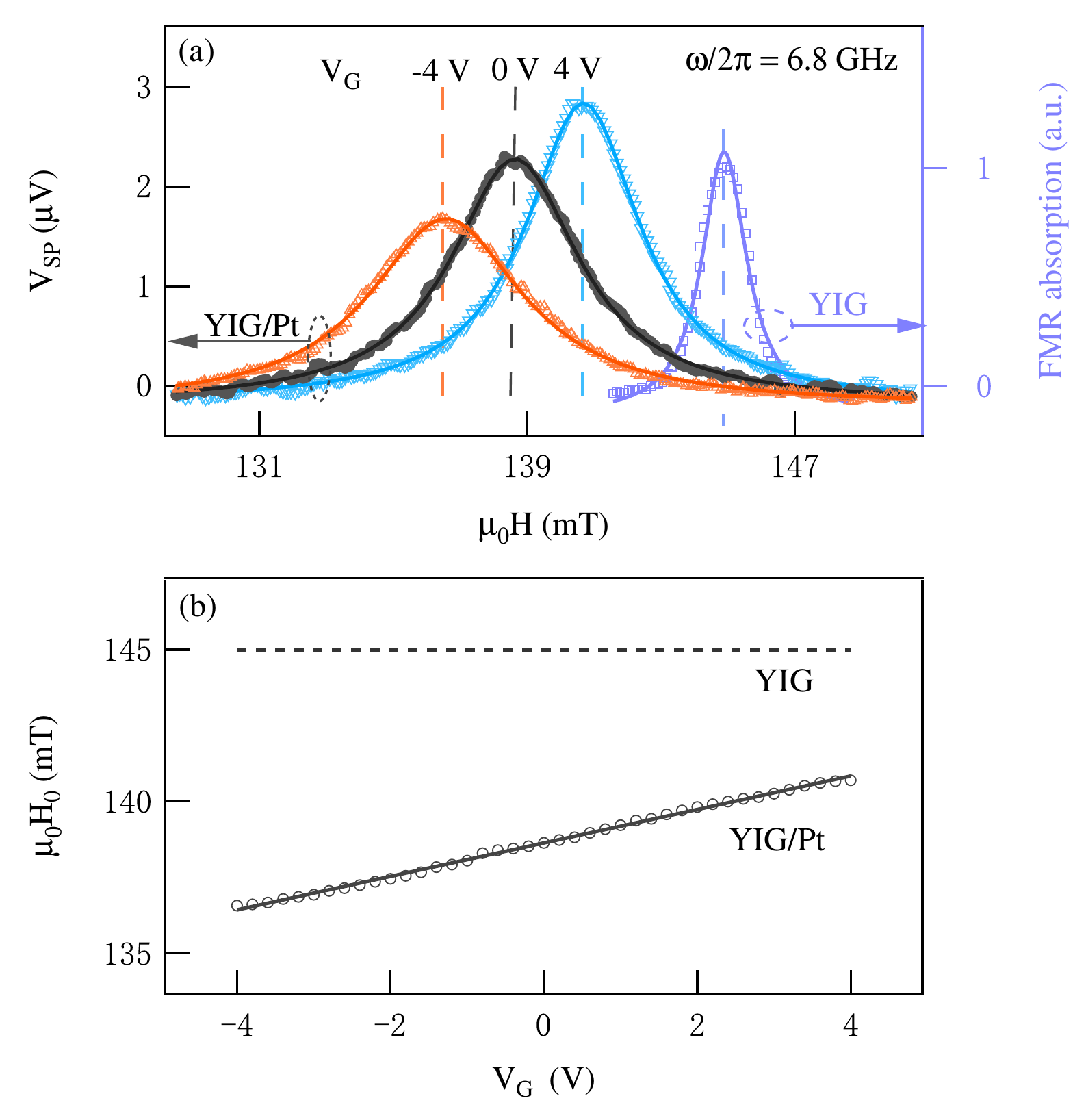}
    \caption{(a) The FMR resonance field $H_{0}$ was shifted to the high field and low field sides for positive and negative gate voltages, respectively.  The FMR absorption in bare YIG is displayed for comparison. (b) The FMR resonance field $H_{0}$ as a function of the gate voltage is summarized and compared to that in a bare YIG layer. Here, the microwave frequency was 6.8 GHz. }
    \label{fig4}
\end{figure}
 
An analogous electrically tunable anomalous Hall effect in Pt was previously reported using an ionic gate technique \cite{Shimizu2013PRL} and may involve similar underlying physics, related to a strongly charge density dependent spin orbit interaction.
However in previous results the gate voltage influence on the resistance and Hall effect was irreversible, whereas the spin mixing conductance control in our work can be repeated many times and may therefore be utilized to control spin current transport in future spintronic devices.
We have performed spin pumping measurements in YIG(20 nm)/Pt($t$ nm) for a variety of Pt thickness $t$\cite{SuppleM}. 
Compared to the large $V_{G}$-tunable effect in the 2.5-nm-thick-Pt sample discussed above,  the spin pumping signal was greatly reduced in a 4-nm-thick-Pt sample and barely observable in a 10-nm-thick-Pt sample. These results further indicate that the effective spin mixing conductance change is due to charge accumulation at the YIG/Pt interface, which is greatly enhanced in the thin Pt samples.
 
The physics underlying the tunable spin mixing conductance and spin Hall angle is the change in carrier density due to the gate voltage, which will also induce a boundary change at the bilayer interface as we highlighted in Fig. \ref{fig1} (c).  Interestingly this also appears to induce an anisotropy change in the YIG/Pt bilayer as a function of the gate voltage.  In Fig. \ref{fig4} (a) we have plotted the FMR signals in the bare YIG layer and in the YIG/Pt bilayer with different gate voltages while the external magnetic field was applied in the film plane.  The vertical dashed lines highlight the resonance positions. 
The large 6.8 mT shift to low fields which is observed in the bilayer device, compared to the bare YIG, is due to the influence of the Pt layer on the boundary conditions of the bare YIG surface.
When we apply a gate voltage the FMR  resonance field $H_{0}$ shifts to higher fields by 2.06 mT for $V_{G}$ = 4.0 V and to lower fields by 2.07 mT for $V_{G}$ = -4.0 V.
The $V_{G}$-dependent $H_{0}$ is summarized in panel (b) and shows nearly a linear dependence.
A more complex anisotropy change was observed in different samples and the mechanism is still an open question for future work, but bears an interesting analogy to the electrical field induced anisotropy changes in systems such as CoFeB/MgO \cite{Wang2012NatMaterials} and CoO/Co \cite{Zhang2015nanoScale, Wang2017nanoScale}.
  
\section{CONCLUSION}
In summary, we report the modulation of the charge carrier density at the interface of a  Y${}_{3}$Fe${}_{5}$O${}_{12}$/Platinum (YIG/Pt) bilayer using an ionic gate technique. 
We electrically detected the ferromagnetic resonance (FMR) at variable gate voltages, observing three major features:
(1) The line width of FMR is controlled by a gate voltage, which indicates that the spin mixing conductance in the bilayer can be tuned; 
(2) The voltage amplitude of spin pumping is strongly dependent on the gate voltage.  To model the voltage change we found that the spin Hall angle in Pt should be a function of the gate voltage; 
(3) The anisotropy change indicates that the boundary conditions at the interface of the bilayer are changed by the gate voltage. 
Thus, we experimentally demonstrated control of spin current due to spin pumping in a YIG/Pt bilayer using a gate voltage. 
This observation may be used to better understand spin transfer at magnetic insulator/heavy metal interfaces. 

\section{ACKNOWLEDGMENTS}
This work is supported by the `National Young  1000 Talents' Program  and  by the National Natural Science Foundation of China (NSFC No. 11774200) grants (Lihui Bai).


\begin{thebibliography}{60}

\bibitem{Kajiwara2010Nature}
Y. Kajiwara, K. Harii, S. Takahashi, J. Ohe, K. Uchida, M. Mizuguchi, H. Umezawa, H. Kawai, K. Ando,K. Takanashi, S. Maekawa, and E. Saitoh,
Transmission of electrical signals by spin-wave interconversion in a magnetic insulator, 
\href{https://doi.org/10.1038/nature08876}{Nature {\bf 464}, 262 (2010)}.

\bibitem{Lu2012PRL}
Lei Lu, Yiyan Sun, Michael Jantz, and Mingzhong Wu
Control of Ferromagnetic Relaxation in Magnetic Thin Films through Thermally Induced Interfacial Spin Transfer,
\href{https://doi.org/10.1103/PhysRevLett.108.257202}{Physical Review Letters, {\bf 108}, 257202 (2012)}. 

\bibitem{Bai2013PRL}
Lihui Bai, P. Hyde, Y. S. Gui, C.-M. Hu, V. Vlaminck, J. E. Pearson, S. D. Bader, and A. Hoffmann,
Universal Method for Separating Spin Pumping from Spin Rectification Voltage of Ferromagnetic Resonance,
\href{https://doi.org/10.1103/PhysRevLett.111.217602}{Physical Review Letters {\bf 111}, 217602 (2013)}.

\bibitem{Hahn2013PRB}
C. Hahn, G. de Loubens, O. Klein, M. Viret, V. V. Naletov, and J. Ben Youssef,
Comparative measurements of inverse spin Hall effects and magnetoresistance in YIG/Pt and YIG/Ta,
\href{https://doi.org/10.1103/PhysRevB.87.174417}{Phys. Rev. B {\bf 87}, 174417, (2013)}.

\bibitem{Sun2013PRL}
Yiyan Sun, Houchen Chang, Michael Kabatek, Young-Yeal Song, Zihui Wang, Michael Jantz, William Schneider, Mingzhong Wu, E. Montoya, B. Kardasz, B. Heinrich, Suzanne G. E. te Velthuis, Helmut Schultheiss, and Axel Hoffmann,
Damping in Yttrium Iron Garnet Nanoscale Films Capped by Platinum,
\href{https://doi.org/10.1103/PhysRevLett.111.106601}{Phys. Rev. Lett. {\bf 111}, 106601, (2013)}.

\bibitem{Castel2014PRB}
V. Castel, N. Vlietstra, B. J. van Wees, and J. Ben Youssef,
Yttrium iron garnet thickness and frequency dependence of the spin-charge current conversion in YIG/Pt systems,
\href{https://doi.org/10.1103/PhysRevB.90.214434}{Phys. Rev. B {\bf 90}, 214434, (2014)}.

\bibitem{Hyde2014PRB}
P. Hyde, Lihui Bai, D. M. J. Kumar, B. W. Southern, C.-M. Hu, S. Y. Huang, B. F. Miao, and C. L. Chien,
Electrical detection of direct and alternating spin current injected from a ferromagnetic insulator into a ferromagnetic metal,
\href{https://doi.org/10.1103/PhysRevB.89.180404}{Phys. Rev. B {\bf 89}, 180404(R), (2014)}.

\bibitem{Haertinger2015PRB}
M. Haertinger, C. H. Back, J. Lotze, M. Weiler, S. Gepr\"{a}gs, H. Huebl, S. T. B. Goennenwein, and G. Woltersdorf,
Spin pumping in YIG/Pt bilayers as a function of layer thickness,
\href{https://doi.org/10.1103/PhysRevB.92.054437}{Phys. Rev. B {\bf 92}, 054437, (2015)}.

\bibitem{Zhou2016PRB}
Hengan Zhou, Xiaolong Fan, Li Ma, Qihan Zhang, Lei Cui, Shiming Zhou, Y. S. Gui, C.-M. Hu, and Desheng Xue,
Spatial symmetry of spin pumping and inverse spin Hall effect in the Pt/Y3Fe5O12 system,
\href{https://doi.org/10.1103/PhysRevB.94.134421}{Phys. Rev. B {\bf 94}, 134421, (2016)}.

\bibitem{Dushenko2016PRL}
S. Dushenko, H. Ago, K. Kawahara, T. Tsuda, S. Kuwabata, T. Takenobu, T. Shinjo, Y. Ando, and M. Shiraishi,
Gate-Tunable Spin-Charge Conversion and the Role of Spin-Orbit Interaction in Graphene,
\href{https://doi.org/10.1103/PhysRevLett.116.166102}{Phys. Rev. Lett. {\bf 116}, 166102, (2016)}.

\bibitem{Wang2016PRL}
Hailong Wang, James Kally, Joon Sue Lee, Tao Liu, Houchen Chang, Danielle Reifsnyder Hickey, K. Andre Mkhoyan, Mingzhong Wu, Anthony Richardella, and Nitin Samarth,
Surface-State-Dominated Spin-Charge Current Conversion in Topological-Insulator-Ferromagnetic-Insulator Heterostructures,
\href{https://doi.org/10.1103/PhysRevLett.117.076601}{Phys. Rev. Lett. {\bf 117}, 076601, (2016)}.

\bibitem{Wesenberg2017NatPhysics}
Devin Wesenberg, Tao Liu, Davor Balzar, Mingzhong Wu and Barry L. Zink,
Long-distance spin transport in a disordered magnetic insulator,
\href{https://doi.org/10.1038/nphys4175}{Nature Physics {\bf 13}, 987, (2017)}.

\bibitem{Kapelrud2017PRB}
A. Kapelrud and A. Brataas,
Spin pumping, dissipation, and direct and alternating inverse spin Hall effects in magnetic-insulator/normal-metal bilayers,
\href{https://doi.org/10.1103/PhysRevB.95.214413}{Phys. Rev. B {\bf 95}, 214413, (2017)}.

\bibitem{Keller2017PRB}
S. Keller, J. Greser, M. R. Schweizer, A. Conca, V. Lauer, C. Dubs, B. Hillebrands, and E. Th. Papaioannou,
Relative weight of the inverse spin-Hall and spin-rectification effects for metallic polycrystalline Py/Pt, epitaxial Fe/Pt, and insulating YIG/Pt bilayers: Angular dependent spin pumping measurements,
\href{https://doi.org/10.1103/PhysRevB.96.024437}{Phys. Rev. B {\bf 96}, 024437, (2017)}.

\bibitem{Tserkovnyak2002PRL}
Yaroslav Tserkovnyak, Arne Brataas, and Gerrit E. W. Bauer,
Enhanced Gilbert Damping in Thin Ferromagnetic Films,
\href{https://doi.org/10.1103/PhysRevLett.88.117601}{Phys. Rev. Lett. {\bf 88}, 117601, (2002)}.

\bibitem{Montoya2014PRL}
Eric Montoya, Bret Heinrich, and Erol Girt,
Quantum Well State Induced Oscillation of Pure Spin Currents in Fe/Au/Pd(001) Systems,
\href{https://doi.org/10.1103/PhysRevLett.113.136601}{Phys. Rev. Lett. {\bf 113}, 136601, (2014)}.

\bibitem{Saitoh2006APL}
E. Saitoh, M. Ueda,  H. Miyajima, and G. Tatara,
Conversion of spin current into charge current at room temperature: Inverse spin-Hall effect,
\href{https://doi.org/10.1063/1.2199473}{Appl. Phys. Lett. {\bf 88}, 182509, (2006)}. 

\bibitem{Tserkovnyak2005RMP}
Yaroslav Tserkovnyak, Arne Brataas, Gerrit E. W. Bauer, and Bertrand I. Halperin,
Nonlocal magnetization dynamics in ferromagnetic heterostructures,
\href{https://doi.org/10.1103/RevModPhys.77.1375}{Rev. Mod. Phys. {\bf 77}, 1375, (2005)}.

\bibitem{Igor2004RMP}
Igor \v{Z}uti\'{c}, Jaroslav Fabian, and S. Das Sarma,
Spintronics: Fundamentals and applications,
\href{https://doi.org/10.1103/RevModPhys.76.323}{Rev. Mod. Phys. {\bf 76}, 323, (2004)}.

\bibitem{Du2013PRL}
C. H. Du, H. L. Wang, Y. Pu, T. L. Meyer, P. M. Woodward, F. Y. Yang, and P. C. Hammel,
Probing the Spin Pumping Mechanism: Exchange Coupling with Exponential Decay in Y$_3$Fe$_5$O$_{12}$/Barrier/Pt Heterostructures,
\href{https://doi.org/10.1103/PhysRevLett.111.247202}{Phys. Rev. Lett. {\bf 111}, 247202, (2013)}.

\bibitem{Wang2014PRL}
Hailong Wang, Chunhui Du, P. Chris Hammel, and Fengyuan Yang,
Antiferromagnonic Spin Transport from Y$_3$Fe$_5$O$_{12}$ into NiO,
\href{https://doi.org/10.1103/PhysRevLett.113.097202}{Phys. Rev. Lett. {\bf 113}, 097202, (2014)}.

\bibitem{Qiu2018NatMaterials}
Zhiyong Qiu, Dazhi Hou, Joseph Barker, Kei Yamamoto, Olena Gomonay and Eiji Saitoh,
Spin colossal magnetoresistance in an antiferromagnetic insulator,
\href{https://doi.org/10.1038/s41563-018-0087-4}{Nature Material {\bf 17}, 577, (2018)}. 

\bibitem{Nguyen2016APL}
Minh-Hai Nguyen, Mengnan Zhao, D. C. Ralph, and R. A. Buhrman,
Enhanced spin Hall torque efficiency in PtAl and PtHf alloys arising from the intrinsic spin Hall effect,
\href{https://doi.org/10.1063/1.4953768}{Appl. Phys. Lett. {\bf 108}, 242407, (2016)}.

\bibitem{Wang2012NatMaterials}
Wei-Gang Wang, Mingen Li, Stephen Hageman and C. L. Chien,
Electric-field-assisted switching in magnetic tunnel junctions,
\href{https://doi.org/10.1038/nmat3171}{Nature Materials {\bf 11}, 64, (2012)}.

\bibitem{Yamada2011Science}
Y. Yamada, K. Ueno, T. Fukumura, H. T. Yuan, H. Shimotani, Y. Iwasa, L. Gu, S. Tsukimoto, Y. Ikuhara, M. Kawasaki,
Electrically Induced Ferromagnetism at Room Temperature in Cobalt-Doped Titanium Dioxide,
\href{https://doi.org/10.1126/science.1202152 }{Science {\bf 332} 1065, (2011)}.

\bibitem{Shimizu2013PRL}
Sunao Shimizu, Kei S. Takahashi, Takafumi Hatano, Masashi Kawasaki, Yoshinori Tokura, and Yoshihiro Iwasa,
Electrically Tunable Anomalous Hall Effect in Pt Thin Films,
\href{https://doi.org/10.1103/PhysRevLett.111.216803}{Phys. Rev. Lett. {\bf 111}, 216803, (2013)}.

\bibitem{Liang2018PRB}
L. Liang, J. Shan, Q. H. Chen, J. M. Lu, G. R. Blake, T. T. M. Palstra, G. E. W. Bauer, B. J. van Wees, and J. T. Ye,
Gate-controlled magnetoresistance of a paramagnetic-insulator/platinum interface,
\href{https://doi.org/10.1103/PhysRevB.98.134402}{Phys. Rev. B {\bf 98}, 134402, (2018)}.

\bibitem{Guan2018AM}
Mengmeng Guan, Lei Wang, Shishun Zhao, Ziyao Zhou, Guohua Dong, Wei Su, Tai Min, Jing Ma, Zhongqiang Hu, Wei Ren, Zuo‐Guang Ye, Ce‐Wen Nan, and Ming Liu,
Ionic Modulation of the Interfacial Magnetism in a Bilayer System Comprising a Heavy Metal and a Magnetic Insulator for Voltage-Tunable Spintronic Devices,
\href{https://doi.org/10.1002/adma.201802902}{Advanced Materials {\bf 30}, 1802902, (2018)}.

\bibitem{SuppleM}
See Supplemental Material at [URL will be  inserted by publisher] for more details on the gate voltage applied to a bare YIG layer and the Pt-thickness dependence in YIG/Pt bilayers.

\bibitem{Weiler2013PRL}
Mathias Weiler, Matthias Althammer, Michael Schreier, Johannes Lotze, Matthias Pernpeintner, Sibylle Meyer, Hans Huebl, Rudolf Gross, Akashdeep Kamra, Jiang Xiao, Yan-Ting Chen, HuJun Jiao, Gerrit E. W. Bauer, and Sebastian T. B. Goennenwein,
Experimental Test of the Spin Mixing Interface Conductivity Concept,
\href{https://doi.org/10.1103/PhysRevLett.111.176601}{Phys. Rev. Lett. {\bf 111}, 176601, (2013)}.

\bibitem{Jiao2013PRL}
HuJun Jiao and Gerrit E. W. Bauer,
Spin Backflow and ac Voltage Generation by Spin Pumping and the Inverse Spin Hall Effect,
\href{https://doi.org/10.1103/PhysRevLett.110.217602}{Phys. Rev. Lett. {\bf 110}, 217602, (2013)}.

\bibitem{Du2014PRApplied}
Chunhui Du, Hailong Wang, Fengyuan Yang, and P. Chris Hammel,
Enhancement of Pure Spin Currents in Spin Pumping Y$_3$Fe$_5$O$_{12}$/Cu/Metal Trilayers through Spin Conductance Matching,
\href{https://doi.org/10.1103/PhysRevApplied.1.044004}{Phys. Rev. Applied {\bf 1}, 044004, (2014)}.

\bibitem{Zhang2015nanoScale}
Kun Zhang,  Yan-ling Cao,  Yue-wen Fang,  Qiang Li,  Jie Zhang,  Chun-gang Duan,  Shi-shen Yan,  Yu-feng Tian,  Rong Huang,  Rong-kun Zheng,  Shi-shou Kang,  Yan-xue Chen,  Guo-lei Liu  and  Liang-mo Mei,  
Electrical control of memristance and magnetoresistance in oxide magnetic tunnel junctions,
\href{https://doi.org/10.1039/C5NR00522A}{Nanoscale {\bf 9}, 6336, (2015)}.

\bibitem{Wang2017nanoScale}
Jing Wang,  Qikun Huang,  Peng Shi,  Kun Zhang,  Yufeng Tian,  Shishen Yan,  Yanxue Chen,  Guolei Liu,  Shishou Kang  and  Liangmo Mei,
Electrically tunable tunneling rectification magnetoresistance in magnetic tunneling junctions with asymmetric barriers,
\href{https://doi.org/10.1039/C7NR04431C}{Nanoscale {\bf 9}, 16073, (2017)}.

\end{thebibliography}
\end{document}